\documentclass[twocolumn,preprintnumbers,prl]{revtex4}
\usepackage{graphicx}
\usepackage{hyperref}

\begin{document}

\preprint{LBNL-53467}

\title{Technicolorful Supersymmetry}

\author{Hitoshi Murayama}
\affiliation{Department of Physics, University of California,
                Berkeley, CA 94720, USA}
\affiliation{Theoretical Physics Group, Lawrence Berkeley National Laboratory,
                Berkeley, CA 94720, USA}
\date{\today}

\begin{abstract}
  Technicolor achieves electroweak symmetry breaking (EWSB) in an
  elegant and natural way, while it suffers from severe model building
  difficulties.  I propose to abandon its secondary goal to eliminate
  scalar bosons in exchange of solving numerous problems using
  supersymmetry.  It helps to understand walking dynamics much better
  with certain exact results.  In the particular model presented here,
  there is no light elementary Higgs boson and the EWSB is fully
  dynamical, hence explaining the hierarchy; There is no alignment
  problem and no light pseudo-Nambu-Goldstone bosons exist; The
  fermion masses are generated by a ultraviolet-complete
  renormalizable extended technicolor sector with techni-GIM mechanism
  and hence the sector is safe from flavor-changing-neutral-current
  constraints; The ``$e^+ e^-$'' production of techni-states in the
  superconformal window is calculable; The electroweak precision
  observables are (un)fortunately not calculable.
\end{abstract}
\maketitle

\section{Introduction}

From the beautiful measurements of the triple gauge boson vertices at
LEP-II \cite{TGV} that agreed with the expectation in the non-abelian
gauge theory, there is little doubt that the $W$ and $Z$ bosons are
indeed gauge bosons of the electroweak $SU(2)_L \times U(1)_Y$
symmetry.  Their finite mass implies that our Universe is filled with
a Bose--Einstein condensate, analogous to Cooper pair condensates in
superconductors.  The analog of the finite penetration length in
superconductors is the finite Compton wavelengths of $W$ and $Z$
bosons, making the interaction short-ranged.  This is the physics of
electroweak symmetry breaking (EWSB).  The energy scale of the
condensate is 250~GeV and understanding the nature of the condensate
is the target of intense experimental activity in this and the next
decade at Tevatron, LHC, and future electron-positron linear collider.

In the Standard Model of particle physics, the EWSB is accomplished by
the condensation of elementary spinless particle called the Higgs
boson.  The model breaks down below TeV scale because the Higgs boson
mass-squared receives a quadratic divergence rendering the theory
unstable.  It is possible to stabilize it using supersymmetry; in this
conventional use of supersymmetry, the Higgs boson is still an
elementary scalar while its mass-squared receives only a logarithmic
divergence to all orders in perturbation theory.  Recently, it has
become a concern that the supersymmetry is getting already somewhat
fine-tuned after the unsuccessful search for superparticles at LEP-II
(see, {\it e.g.}\/, \cite{Barbieri:2000gf}), sometimes referred to as
the ``little hierarchy'' problem.

Technicolor theories, on the other hand, attempt to describe the EWSB
much more analogously to the superconductivity.  The condensate is not
elementary, but rather a pair of fermions.  However, the lack of large
Fermi surface in the Dirac sea in the relativistic theory of fermions
forces us to introduce a new {\it strong}\/ interaction.  This point
immediately poses difficulty in model building, as the models rely on
non-perturbative strong dynamics of gauge theories.  In addition to
this technical difficulty, there had been numerous other
phenomenological difficulties.  I refer readers to an excellent review
for model-building attempts and an up-to-date list of references
\cite{Hill:2002ap}.

The main difficulties in the original QCD-like technicolor models are
as follows:
\begin{itemize}
\item Technicolor itself does not provide a mechanism of generating
  masses of quarks and leptons.  It has to be augmented by an extended
  technicolor (ETC) sector.  The large top quark mass implies the ETC
  scale is as low as a few TeV, while the flavor-changing neutral
  current (FCNC) constraints require the ETC scale to be above 10--100
  TeV.  Some ETC models even suffer from proton instability.
\item Most models have the alignment problem, namely that there are
  degenerate ground states in technicolor theory some of which do not
  successfully achieve the EWSB.  Even if the correct ground state is
  chosen, there are light pseudo-Nambu-Goldstone bosons (PNGB) which
  are experimentally ruled out.
\item Certain precisely measured electroweak observables measured are
  predicted to be out of the allowed ranges.
\end{itemize}

There had been efforts to solve these problems using non-QCD-like
dynamics, especially using the ``walking'' theories
\cite{Holdom:1981rm,Yamawaki:1985zg,Appelquist:an} (see also
\cite{Hill:2002ap} and references therein).  In these theories, it is
assumed that there are large anomalous dimensions that enhances the
ETC interactions and solve the phenomenological problems listed above.
Other possible directions include topcolor-assisted technicolor, extra
dimensions, topseesaw, etc \cite{other}.

Apart from the EWSB itself, there has been another motivation for
technicolor models: to eliminate spinless bosons from the theory.  The
argument is that we have not yet seen any spinless elementary bosons
in Nature, and their masses suffer from the quadratic divergences
which render the theory unstable and fine-tuned.  I consider this a
secondary goal less important than the primary goal of the elegant
EWSB.  If the secondary goal is abandoned, it is attractive to
consider the possibility that the technicolor model is supersymmetric.
It necessarily introduces spinless bosons into the theory, while it
will not cause dangerous quadratic divergences.  On the other hand,
non-perturbative dynamics of supersymmetric gauge theories are much
better understood than the non-supersymmetric counter parts which
provides better tools in model building (see
\cite{Intriligator:1995au} for a review).  In fact, I will make use of
the superconformal dynamics which allows exact predictions on the
anomalous dimensions.  Moreover, the presence of spinless bosons will
provide simple solutions to the alignment problem and the successful
construction of the ETC sector as I present below.

Here are the relevant energy scales in the particular model presented
below:
\begin{itemize}
\item $m_{SUSY} \sim 0.2$~TeV.
\item $\Lambda_{TC} \sim 2$~TeV.
\item $\Lambda_{ETC} \sim 2$~TeV.
\end{itemize}
The EWSB is solely due to the strong dynamics of technicolor gauge
interaction, and supersymmetry breaking is considered a small
perturbation.  The main advantage of using supersymmetry is to have
clear predictions on the dynamics.  Certain quantities are exactly
calculable despite strong dynamics of the theory.  Despite a
relatively low ETC scale, supersymmetry allows a renormalizable
implementation of the techni-GIM mechanism \cite{Randall:1992vt} to
avoid the FCNC problems.  

Earlier attempts to make use of supersymmetry in the context of
technicolor have assumed $m_{SUSY} > \Lambda_{TC}$, and moreover were
made before the dynamics of supersymmetric gauge theories were
understood \cite{Dine:1990jd,other}.  The effect of soft supersymmetry
breaking on dynamics had been worked out only relatively recently
\cite{Luty:1999qc}.  There is one important exception that used
$m_{SUSY} < \Lambda_{TC}$ and made use of supersymmetric dynamics by
Luty, Terning, and Grant \cite{Luty:2000fj} which, however, needed a
light Higgs boson.  The present work makes use of these developments.

\section{Technicolor}

The simplest choice of the technicolor group is $SU(2)_{TC}$ with the
following particle content of chiral superfields (techniquarks)
\begin{equation}
  T({\bf 2}, {\bf 2}, 0), \quad t_+ ({\bf 2}, {\bf 1}, +\frac{1}{2}),
  \quad t_- ({\bf 2}, {\bf 1}, -\frac{1}{2})
\end{equation}
where the quantum numbers are shown under the gauge group $SU(2)_{TC}
\times SU(2)_L \times U(1)_Y$.  This is the same quantum number
assignment as the minimal technicolor model in the non-supersymmetric
case.  The dynamics of technicolor treats two components of $T$,
$t_+$, and $t_-$ all equal, and hence there is an $SU(4)$ global
symmetry.  The model described here was studied by Luty, Terning, and
Grant in \cite{Luty:2000fj}.

The low-energy effective theory of this supersymmetric gauge theory is
known, and is described by the meson composites made of techniquarks
\cite{Intriligator:1995au}.  It is known to lead to the so-called
quantum modified constraint:
\begin{equation}
  M_+ M_- - M_S M_s = \Lambda^4,
\end{equation}
where the mesons are defined by
\begin{equation}
  M_\pm = (T t_\pm), \qquad M_S = (T T), \qquad M_s = (t_+ t_-).
\end{equation}
The contraction of technicolor indices is understood in each
parentheses.  At this point the model suffers from the alignment
problem just like in the non-supersymmetric model.  It is not clear
how the desired ground state with expectation values in electroweak
doublet composites $M_+$ and $M_-$ is chosen over that with singlets
$M_S$ and $M_s$ that do not break the electroweak gauge group at all.
It is often assumed that the higher the remaining symmetry the lower
the energy is; if so, it is more likely that the ground state does not
achieve the EWSB.  (Here, $\Lambda$ is the holormorphic dynamical
scale, which is different from the strong scale of technicolor theory.
We will come back to this point later.)

In this model, it is very simple to solve the alignment problem.  I
introduce two singlet fields $S$ and $s$, and a superpotential
\begin{equation}
  W = S (T T) + s (t_+ t_-).
\end{equation}
This superpotential is renormalizable and hence is ultraviolet
complete.  Once the technicolor interaction becomes strong at the
scale $\Lambda$, the superpotential turns into mass terms of meson
composites together with the introduced singlets of the order of
$\Lambda$.  The superpotential does not allow the $M_S = (T T)$ and
$M_s = (t_+ t_-)$ mesons to acquire expectation values.  This way, the
desired ground state with the EWSB
\begin{equation}
  \langle M_+ \rangle = \langle M_- \rangle = \Lambda^2
\end{equation}
is uniquely chosen together with the $D$-term potential, solving the
alignment problem.  By further making the soft masses for $t_+$ and
$t_-$ different in the ultraviolet, one can achieve different VEVs for
$M_\pm$ as well ({\it i.e.}\/, $\tan \beta \neq 1$).

Note that the dynamics has the custodial $SU(2)$ symmetry and hence
leads to $\rho =1$ naturally.  Out of the $SU(4)\simeq SO(6)$ global
symmetry, the superpotential breaks it to $SO(4) \simeq SU(2)_L \times
SU(2)_R$.  $U(1)_Y$ is embedded into $SU(2)_R$.  Therefore, the
custodial $SU(2)$ symmetry is broken only by the $U(1)_Y$ interaction
and fermion masses (see the next section), in complete analogy to the
minimal Standard Model.

The effect of soft supersymmetry breaking in this model had been
worked out by Luty and Rattazzi \cite{Luty:1999qc}.  Out of five
chiral superfields that survive the constraint, the five imaginary
components acquire positive mass squared of $O(m_{SUSY}^2)$ while the
five real components remain massless.  The massless ones are the
Nambu--Goldstone bosons of the spontaneously broken symmetry
$SU(4)/Sp(2) = SO(6)/SO(5)$.  Together with the terms from the
superpotential, two full chiral multiplets become massive, leaving
only three real and imaginary components left.  The three massless
imaginary components are eaten by the $W$ and $Z$ bosons as a
consequence of the EWSB, while the real parts remain with
supersymmetry-breaking scale.  They are no Higgs bosons, however, as
the Higgs boson is completely eliminated by the constraint.  They are
rather the analog of $H^0$ and $H^\pm$ in two-doublet Higgs models
while they lack scalar-scalar-vector vertices.

Therefore the dynamical EWSB works successfully in this model.

\section{ETC}

In the original QCD-like technicolor, the fermion masses are obtained
through dimension-six four-fermion interactions giving
\begin{equation}
  m_f \sim \frac{1}{(4\pi)^2} \frac{\Lambda_{TC}^3}{\Lambda_{ETC}^2}.
\end{equation}
The trouble is that one also expects flavor-changing neutral currents
from the ETC boson exchange suppressed by the same power,
$1/\Lambda_{ETC}^2$.  This causes the well-known dilemma; a large
enough fermion mass, especially that for the top quark, is
incompatible with the apparent lack of flavor-changing neutral current
processes.  

The fermion masses are obtained through dimension-{\it five}\/
superpotential terms
\begin{eqnarray}
  \lefteqn{
    W_{ETC} = h_u^{ij} \frac{1}{\Lambda_{ETC}} (T t_+) Q_i u_j}\\
  & & + h_d^{ij} \frac{1}{\Lambda_{ETC}} (T t_-) Q_i d_j
  + h_l^{ij} \frac{1}{\Lambda_{ETC}} (T t_-) L_i e_j.
\end{eqnarray}

In the supersymmetric case based on the model in the previous section,
the ETC scale cannot be raised high \cite{Luty}.  The naive
dimensional analysis is used to relate various scales in the problem
\cite{Luty:1997fk,Cohen:1997rt}, as the holormorphy in supersymmetry
is not powerful enough to constrain the K\"ahler potential.  It
suggests that the Lagrangian for the meson composites is given by
\begin{eqnarray}
  \lefteqn{
    L = \frac{1}{(4\pi)^2} \left[
      \int d^4 \theta \hat{M}_\pm^\dagger \hat{M}_\pm \right. } \nonumber
    \\
    & & \left.
      + \int d^2 \theta \hat{X} (\hat{M}_+ \hat{M}_- - \Lambda_{TC}^2) 
      + \Lambda_{TC} \hat{M}_- \frac{Q U}{\Lambda_{ETC}} \right].
\end{eqnarray}
The meson composite fields $\hat{M}_\pm$ have different normalization
from $M_\pm$ as will be determined shortly below.  It gives $m_W^2
\simeq g^2 v^2$ with $v \sim \Lambda_{TC}/4\pi$.  This result is
analogous to the result in non-supersymmetric QCD $f_\pi \sim
\Lambda/(4\pi)$.  $\hat{X}$ is a Lagrange multiplier field to enforce
the quantum modified constraint.  The last term is the desired ETC
interaction responsible for the top quark mass.  It leads to the mass
term
\begin{equation}
  \frac{1}{(4\pi)^2} \frac{\Lambda_{TC}^2}{\Lambda_{ETC}} Q U
\end{equation}
which implies
\begin{equation}
  m_f \sim \frac{1}{(4\pi)^2} \frac{\Lambda_{TC}^2}{\Lambda_{ETC}}
  \simeq \frac{v^2}{\Lambda_{ETC}}.
\end{equation}
In order to reproduce the top quark mass, I need $\Lambda_{ETC} \sim
v$.  Hence the presumed high ETC scale is brought down to the
electroweak scale and it does not fulfill the goal.  It is also
important to note that the strong scale $\Lambda_{TC} \simeq 4\pi v$
is different from Seiberg's holomorphic scale $\Lambda \sim v$ and the
meson composites are related by $\hat{M}_\pm \sim 4\pi M_\pm
/\Lambda$.

Another quantity that cannot be calculated rigorously is the
contribution to the oblique electroweak parameters such as
Peskin--Takeuchi $S$ and $T$ \cite{Peskin:1990zt}.

\section{Walking}

As in the non-supersymmetric case, walking dynamics can be employed to
enhance the ETC interaction to raise the ETC scale and suppress the
possible FCNC effects.  The main benefit of the supersymmetry is to
allow me to calculate the anomalous dimension factors exactly.  

In supersymmetric $SU(N_c)$ QCD for $\frac{3}{2}N_c < N_f < 3 N_c$,
the theory is in superconformal phase \cite{Intriligator:1995au}.  It
has a dual magnetic description in terms of the dual gauge group
$SU(N_f - N_c)$.  The important point is that the wave function
renormalization factor for quark fields is given exactly in the
infrared as
\begin{equation}
  Z = \left( \frac{\mu}{\Lambda} \right)^{(3N_c - N_f)/N_f},
\end{equation}
where $\mu$ ($\Lambda$) is the infrared (ultraviolet) scale.  In
particular, the case $N_c=2$ and $N_f =4$ will be used in the next
section, giving
\begin{equation}
  Z = \left( \frac{\mu}{\Lambda} \right)^{1/2}.
\end{equation}
Note that the suppressed wave function renormalization factor $Z$ at
lower energies {\it enhances}\/ the couplings of techniquarks.



\section{An Overkill ETC Model}

The model introduces four additional chiral multiplets with the same
quantum numbers as the Higgs doublets $H_u$ and $H_d$ in the Minimal
Supersymmetric Standard Model (MSSM).  They are singlets under the
technicolor gauge group.
\begin{equation}
  \Phi_u, \Phi'_u ({\bf 1}, {\bf 2}, +\frac{1}{2}), \quad
  \Phi_d, \Phi'_d ({\bf 1}, {\bf 2}, -\frac{1}{2}).
\end{equation}
They have the superpotential
\begin{eqnarray}
  W_{ETC} &=& \Lambda_{ETC} (\Phi_u \Phi'_d + \Phi'_u \Phi_d) + \Phi'_u (T t_-)
  + \Phi_d' (T t_+) \nonumber \\
  &&+ h_u^{ij} Q_i u_j \Phi_u + h_d^{ij} Q_i d_j \Phi_d
  + h_l^{ij} L_i e_j \Phi_d.
\end{eqnarray}
The only flavor-violating couplings are the Yukawa couplings
$h_u^{ij}$, $h_{d}^{ij}$, $h_l^{ij}$, and hence the model realizes the
requirement of the techni-GIM mechanism.  By integrating out the
massive $\Phi_u$ and $\Phi_d$, I find
\begin{equation}
  W_{ETC}^{\it eff} = \frac{1}{\Lambda_{ETC}} 
  [(h_d^{ij} Q_i d_j + h_l^{ij} L_i e_j)(T t_-)
  + h_u^{ij} Q_i u_j (T t_+)].
\end{equation}
It has all the operators needed to generate fermion masses.  This
model does not lead to any phenomenologically problematic
flavor-changing effects from the ETC operators.

As it stands now, combined with the $SU(2)$ technicolor model with two
flavors in the earlier section, the ETC scale cannot be raised high,
$\Lambda_{ETC} \sim v$.  Therefore this overkill model actually
predicts $\Phi_{u,d}$ are light Higgs doublets and $\Lambda_{ETC}$ is
nothing but the $\mu$ parameter \cite{Luty:2000fj}.  The rest of the
discussion is how the superconformal theory can be used to raise
$\Lambda_{ETC}$ and hence eliminate light Higgs from the spectrum.

Let me take the same $SU(2)$ technicolor model as before, but
introduce two more flavors (four more doublets) to the model.  The
theory becomes strong at a scale $\Lambda_4$ (the subscript stands for
four flavors) which is taken to be much higher than the ETC scale.
Below $\Lambda_4$, the $O(1)$ Yukawa interactions between the $\Phi$
doublets and the techniquarks $\Phi'_u (T t_-) + \Phi_d' (T t_+)$ are
enhanced down to the mass $\Lambda_{ETC}$ of $\Phi$ doublets as
\begin{equation}
    \left( \frac{\Lambda_4}{\Lambda_{ETC}} \right)^{1/2}.
\end{equation}
However, a too-large Yukawa interaction is likely to upset the
delicate conformal dynamics.  It is probably wise to restrict the
growth of the Yukawa coupling to values less than $4\pi$, which I take
as the maximum allowed value of the Yukawa coupling at the ETC scale.
After integrating out the ETC doublets $\Phi_{u,d}$, the effective
interaction is further enhanced by an additional factor
\begin{equation}
    \left( \frac{\Lambda_{ETC}}{\Lambda_{TC}} \right)^{1/2}.
\end{equation}
In the end, an overall enhancement factor of
\begin{equation}
  4\pi \left( \frac{\Lambda_{ETC}}{\Lambda_{TC}} \right)^{1/2}
\end{equation}
can be obtained relative to the previous case.  The technicolor model
reduces to that of the two doublets discussed earlier by adding a mass
term to the extra two doublets $m_{3,4}$ which I take to be a common
mass $m$.  Because the dynamics is already strong, $\Lambda_{TC}$ is
expected to be at the physical mass of the extra doublets
$\Lambda_{TC} \sim m$, leading to the quantum modified constraint and
hence the EWSB.  The fermion mass is then enhanced to
\begin{equation}
  m_f \sim \frac{v^2}{\Lambda_{ETC}} 4\pi \left(
  \frac{\Lambda_{ETC}}{\Lambda_{TC}} \right)^{1/2}, 
\end{equation}
and in order to obtain $m_f \sim v$ (top quark), I need
\begin{equation}
  \Lambda_{ETC} \sim \Lambda_{TC} \sim \frac{\Lambda_4}{(4\pi)^2}.
\end{equation}

Therefore, this model eliminates Higgs doublets at scale $v$ and
additional degrees of freedom such as $\Phi_{u,d}$ are pushed up to
$\Lambda_{TC}$.

\section{Phenomenology}

Phenomenology of the model presented is quite rich.  Below TeV, the
model looks very much like any supersymmetric models except that there
is no light Higgs boson.  There are analogs of heavy Higgses $H^\pm$
and $H^0$ but not $A^0$, even though they are not really Higgs bosons
and there is no vector-vector-scalar vertex.  On the other hand, the
absence of light Higgs implies that the $WW$ scattering amplitude
grows and is unitarized only above TeV.  It may even lead to some
techni-resonances.  If the ETC sector is not an overkill, there may be
small deviations from the Standard Model in $K$- and $B$-physics, and
some lepton-flavor violating signals.

The most striking prediction of this model is the presence of
superconformal dynamics above $\Lambda_{TC}$.  In particular, the
``$e^+ e^-$'' cross section for producing techni-states can be
predicted exactly \cite{deGouvea:1998ft} even though the $S$-matrix
elements are ill-defined in conformal theories.

It is important to note that the usual flavor-changing and CP problems
in supersymmetry exist also in this framework.  The flavor-changing
problems need to be solved by flavor-blind supersymmetry breaking
mechanisms such as gauge mediation \cite{Dine:1995ag}, anomaly
mediation \cite{AMSB} (supplemented by $U(1)$ $D$-terms to make it
viable \cite{Dterm}), or gaugino mediation \cite{gaugino}.  The
hierarchy is stable because of the technicolor rather than
supersymmetry, and it may be possible to push the supersymmetry
breaking scale to the technicolor scale or even beyond, while
maintaining the natural hierarchy of the electroweak scale.  If so,
the flavor-changing problems of supersymmetry may be suppressed partly
by decoupling, and also the little hierarchy problem can be
ameliorated.

The overkill ETC model above should be considered only a toy model as
it does not explain the origin of flavor.  It is conceivable that this
framework can be combined with Froggatt--Nielsen mechnanism
\cite{Froggatt:1978nt} at the ETC scale.  The effect of $O(1)$ Yukawa
couplings on supersymmetry breaking parameters can still be made
immune from FCNC effects using the anomaly mediation.

\section{Open Questions}

The result in this letter is a good start, but poses many open
questions, some theoretical, some phenomenological, and some
aesthetical.  Here is an incomplete list:
\begin{itemize}
\item Is there a way to reliably calculate the electroweak precision
  observables such as $S$ and $T$ parameters?
\item Is there a way to reliably relate the techni-pion decay constant
  to $\Lambda_{TC}$?
\item How far can the supersymmetry breaking scale be pushed up?
  Because the hierarchy is explained by technicolor rather than
  supersymmetry, it may be pushed up to $\Lambda_{TC}$, ameliorating
  the ``little hierarchy'' problem.  If it is pushed even beyond
  $\Lambda_{TC}$, at some point the theoretical control thanks to
  supersymmetry is lost, but maybe reasonable extrapolation on
  dynamics is possible.
\item Can one push up the ETC scale further? 
\item How is the coincidence understood that $m_{SUSY}$,
  $\Lambda_{ETC}$ and $\Lambda_{TC}$ (determined by the masses of
  extra doublets) are not very different?  This is the analog of the
  $\mu$-problem in the MSSM.
\item The true ETC sector is supposed to explain the origin of flavor.
  Can a realistic model of flavor be implemented within this
  framework?
\item Is there a grand-unifiable model?
\end{itemize}

\section{Conclusion}

I showed that the technicolor models with supersymmetry retain the
beauty of the original technicolor model in explaining the EWSB
dynamically with a natural origin for the hierarchy.  It abandons one
of the conventional motivation for technicolor to eliminate spinless
bosons from the theory. On the other hand it solves many of the
problems that have plagued technicolor models without supersymmetry by
walking dynamics under good theoretical control.  There is no
alignment problem and no light pseudo-Nambu-Goldstone bosons exist.
It is easy to generate fermion masses while suppressing the
flavor-changing effects.  In an overkill model presented in this
letter, the fermion masses are generated by a ultraviolet-complete
renormalizable extended technicolor sector with techni-GIM mechanism
and hence the sector is safe from flavor-changing neutral currents
effects.  The electroweak precision observables are not calculable and
I cannot assess the phenomenological constraints at the moment.  On
the other hand the ``$e^+ e^-$'' production of techni-sector can be
calculable despite its walking dynamics.  Given the model building is
relatively simple, there may well be models consistent with grand
unification.

\begin{acknowledgments}
  I thank the organizers of European Physical Society, International
  Europhysics Conference on High Energy Physics, July 17th--23rd,
  2003, in Aachen, Germany, where this work was conceived and
  completed.  I especially thank Markus Luty, who pointed out a
  serious mistake in the first version of the paper, and explained the
  naive dimensional analysis patiently to me.  I also thank Alex Kagan
  and Ken Lane for comments.  This work was supported in part by the
  Director, Office of Science, Office of High Energy and Nuclear
  Physics, of the U.S.  Department of Energy under Contract
  DE-AC03-76SF00098, and in part by the National Science Foundation
  under grant PHY-00-98840.
\end{acknowledgments}
                                

\end{document}